\documentclass[10pt, conference, letterpaper]{IEEEtran}
\makeatletter
\def\ps@headings{%
\def\@oddhead{\mbox{}\scriptsize\rightmark \hfil \thepage}%
\def\@evenhead{\scriptsize\thepage \hfil \leftmark\mbox{}}%
\def\@oddfoot{}%
\def\@evenfoot{}}
\makeatother
\ifCLASSINFOpdf
   \usepackage[pdftex]{graphicx}
\else
   \usepackage[dvips]{graphicx}
\fi

\DeclareRobustCommand*{\IEEEauthorrefmark}[1]{\raisebox{0pt}[0pt][0pt]{\textsuperscript{\footnotesize #1}}}

\usepackage{caption}
\usepackage{array}
\usepackage{soul}
\usepackage{booktabs}
\usepackage{lipsum}
\usepackage{comment}
\usepackage{color}

\usepackage{subcaption}

\usepackage{pifont}
\usepackage{algorithm}
\usepackage{algpseudocode}

\newcommand{\ls}[1]  
  {\dimen0=\fontdimen6\the=#1\dimen0
   \advance\lineskip.5\fontdimen5\the\lineskip-\dimen0
   \lineskiplimit=.9\lineskip
   \baselineskip=\lineskip
   \advance\baselineskip\dimen0
   \normallineskip\lineskip
   \normallineskiplimit\lineskiplimit
   \normalbaselineskip\baselineskip
   \ignorespaces
  }

\ls{1}

\begin{document}

%
\title{Characterizing Location-based Mobile Tracking in Mobile Ad Networks}

\author{\IEEEauthorblockN{
 Boyang Hu\IEEEauthorrefmark{1},   
 Qicheng Lin\IEEEauthorrefmark{1},   
  Yao Zheng\IEEEauthorrefmark{2},      
 Qiben Yan\IEEEauthorrefmark{1},    
 Matthew Troglia\IEEEauthorrefmark{2},		
 Qingyang Wang\IEEEauthorrefmark{3}
 }                                     
 \IEEEauthorblockA{\IEEEauthorrefmark{1}
 The Department of Computer Science \& Engineering, University of Nebraska-Lincoln, Lincoln, NE, 68588, USA}
 \IEEEauthorblockA{\IEEEauthorrefmark{2}
 The Department of Electrical Engineering, University of Hawai`i at M\={a}noa, Honolulu, HI, 96822, USA}
 \IEEEauthorblockA{\IEEEauthorrefmark{3}
 The Department of Computer Science \& Engineering, Louisiana State University, Baton Rouge, LA, 70803, USA}
 \vspace{-25pt}
 }



\maketitle

\begin{abstract}
Mobile apps nowadays are often packaged with third-party ad libraries to monetize user data. Many mobile ad networks exploit these mobile apps to extract sensitive real-time geographical data about the users for location-based targeted advertising. However, the massive collection of sensitive information by the ad networks has raised serious privacy concerns. Unfortunately, the extent and granularity of private data collection of the location-based ad networks remain obscure. In this work, we present a mobile tracking measurement study to characterize the severity and significance of location-based private data collection in mobile ad networks, by using an automated fine-grained data collection instrument running across different geographical areas. We perform extensive threat assessments for different ad networks using 1,100  popular apps running across 10 different cities. This study discovers that the number of location-based ads tend to be positively correlated with the population density of locations, ad networks' data collection behaviors 
differ across different locations, and most ad networks are capable of collecting precise location data. Detailed analysis further reveals the significant impact of geolocation on the tracking behavior of targeted ads, and a noteworthy security concern for advertising organizations to aggregate different types of private user data across multiple apps for a better targeted ad experience. 
\end{abstract}

{\smallskip \IEEEkeywords Ad networks; data privacy; location-based advertising; network traffic analysis.}

\section{Introduction}
Digital advertising market has changed dramatically since the invention of mobile devices. According to Statista \cite{statistaDigitalAdvertising}, while desktop ad spending remains roughly the same, mobile ad spending has grown from 1.57 million U.S. dollars in 2011 to 50.84 billion dollars in 2017. The tremendous growth is in sync with the increasing popularity of mobile devices and apps. Based on the Flurry Analytics data \cite{flurrydata}, an average U.S. consumer nowadays spends five hours a day on mobile apps. Such extended mobile screen time allows ad networks to collect detailed user profiles via ads-enabled mobile apps. In particular, mobile advertising networks have been known to collect data based on users' locations, which allows advertisers to launch targeted advertising campaigns. 
With a controlled ad campaign experiment, a recent study \cite{thinkwithgoogle} shows that location-based ad campaign increases daily active users by 85\%.

However, despite its popularity, the privacy implications of location-based data collection require further scrutiny. Existing works on mobile privacy have focused on longitudinal studying, i.e., how privacy leakage changes over time, using methods such as static analysis~\cite{egele2011pios}, dynamic analysis~\cite{enck2014taintdroid}, network traffic analysis~\cite{razaghpanah2018apps}, and hybrid approaches~\cite{ren2018bug}. But transverse studies regarding location-based data collection, i.e., \emph{how privacy leakage changes across locations}, have not been well investigated.

Compared to longitudinal studies, transverse studies collect data samples across a wide range of locations at a specific point in time. Such methodology poses a challenge for the large-scale measurement. To obtain unbiased results, the researchers must collect mobile traffic at multiple locations and try to eliminate ``time" as an independent variable. Not only does such an experiment require a large number of devices and location samples, but it also relies on effective location spoofing to bypass the ad networks' location verifications. Naive approaches that spoof GPS signals can be easily detected by cross-referencing the GPS coordinates with the devices' network profiles or users' regular activities.

To overcome the challenges, we construct an automated fine-grained data collection instrument running across different geographical areas. We identify the \emph{hot zones} and \emph{cold zones} for mobile privacy study using the physical locations of the advertisers whose websites contain online tracking contents. Our intuition is that mobile tracking originates from web tracking, and therefore the advertisers engaging in web tracking also likely adopt mobile tracking. To bypass location verification, we consistently spoof the GPS coordinates, network profile, and user activities, to let them appear coherent in the eyes of ad networks.

Our instrument also allows us to understand how mobile ad networks aggregate information across apps, by running multiple apps packed with the same ad library. With the lack of tracking cookies in mobile apps, ad networks incorporate ad libraries in different mobile apps which request different sets of permissions. 
By linking the permission profiles of different active apps at different locations to the same ad network, we could evaluate the extent to which the ad networks fuse users' information for targeted advertising.


Based on our extensive measurement with real-world apps, we make a number of interesting discoveries: 1) mobile web ads and mobile in-app ads contact a similar set of popular third-party domains; 2) although the mobile ad network traffic are relatively secure, the low adoption of HTTPs at the advertisers' side still lead to the leakage of private information; 3) different ad networks present different private information collection behaviors across different locations, some of which reveal special interests in collecting particular types of private information; 4) most ad networks can infer users' precise locations even without collecting fine-grained GPS coordinates.

This paper makes the following contributions.

\begin{itemize}
\item We design an efficient privacy leakage measurement system to characterize fine-grained location-based mobile tracking. The system can adjust the GPS locations and network profiles, conduct traffic collection, and perform detailed traffic analysis. 

\item We develop novel domain classification mechanisms to accurately classify the collected domains into ad network domains, advertiser domains, and location-based ad domains.

\item We identify the private data collection behaviors of ad networks at the organization level. We find that there is an alarmingly comprehensive set of users' private information that the ad network organizations can collect by aggregating data from multiple apps.

\item We expose ad networks' information collection behaviors across different locations. Our findings suggest that ad networks manifest different private information collection behaviors at different locations. 
Location leakage by ad networks is particularly disconcerting, as most ad networks can either collect or infer precise locations.

\end{itemize}

\section{Related Work}



The existing privacy research on the mobile ad networks mainly focuses on the malicious uses of advertising contents, which include malicious adSDKs and malicious ad creative. Earlier studies suggested that adSDKs often have poor security and  exhibit fraudulent behaviors~\cite{crussellMAdFraudInvestigatingAd2014}.
Researchers have raised concerns of malicious advertisers recently~\cite{sonWhatMobileAds2016, vines2017exploring}, who can obscure the apps' background to hide malicious activities. In response, ad networks are rapidly improving their screening process to filter out malicious ads and require a minimum number of targeted audiences to prevent individual targeting~\cite{vines2017exploring}.

Demetriou et al. \cite{demetriou2016free} present the first measurement system to reveal the potential risk of ad libraries in mobile apps. Recently, researchers have discovered that the third-party ad libraries in mobile apps misuse their inherited permission and access rights to learn and track users' private information without explicit consent~\cite{razaghpanah2018apps, ren2018bug}. 
Both static and dynamic analyses tools have been developed to detect privacy leakage in mobile apps. 


\noindent \textbf{Static analysis approaches.}
Static analysis is largely scalable and has a low overhead to perform, and it identifies potential privacy leakage through application code analysis. Static analysis of application binaries has been used to detect malicious data flows \cite{egele2011pios}, malware classification \cite{grosse2016adversarial}, 
and user activity analysis \cite{zheng2012smartdroid}.
The changes across different versions of ad libraries~\cite{backes2016reliable} have made the mobile systems more vulnerable because of the adjustments in permission requests across platform/app versions. 

\noindent \textbf{Dynamic analysis approaches.}
Existing studies have provided useful tools to identify the misuse of privacy data through dynamic tainting analysis~\cite{enck2014taintdroid}. The location leakage through location-based services (LBS) has been analyzed~\cite{gibler2012androidleaks,fruchter2015variations}. 
In this paper, we analyze apps across different cities in United States to understand the behaviors of mobile ad networks across different locations. We also consider cross-application privacy leakage by aggregating the collected private information from the same ad domain across multiple apps. 
For improving the coverage of dynamic analysis, researchers have developed ``UI Monkeys" to automate the input generation. Customizable tools like Android Studio's Monkey\footnote{https://developer.android.com/studio/test/monkey} and Appium\footnote{http://appium.io/docs/en/about-appium/intro/}  allow researchers to provide a customized simulation of app interactions. 

\section{Background}

\begin{figure}[t]
\centering
\includegraphics[width=0.9\linewidth]{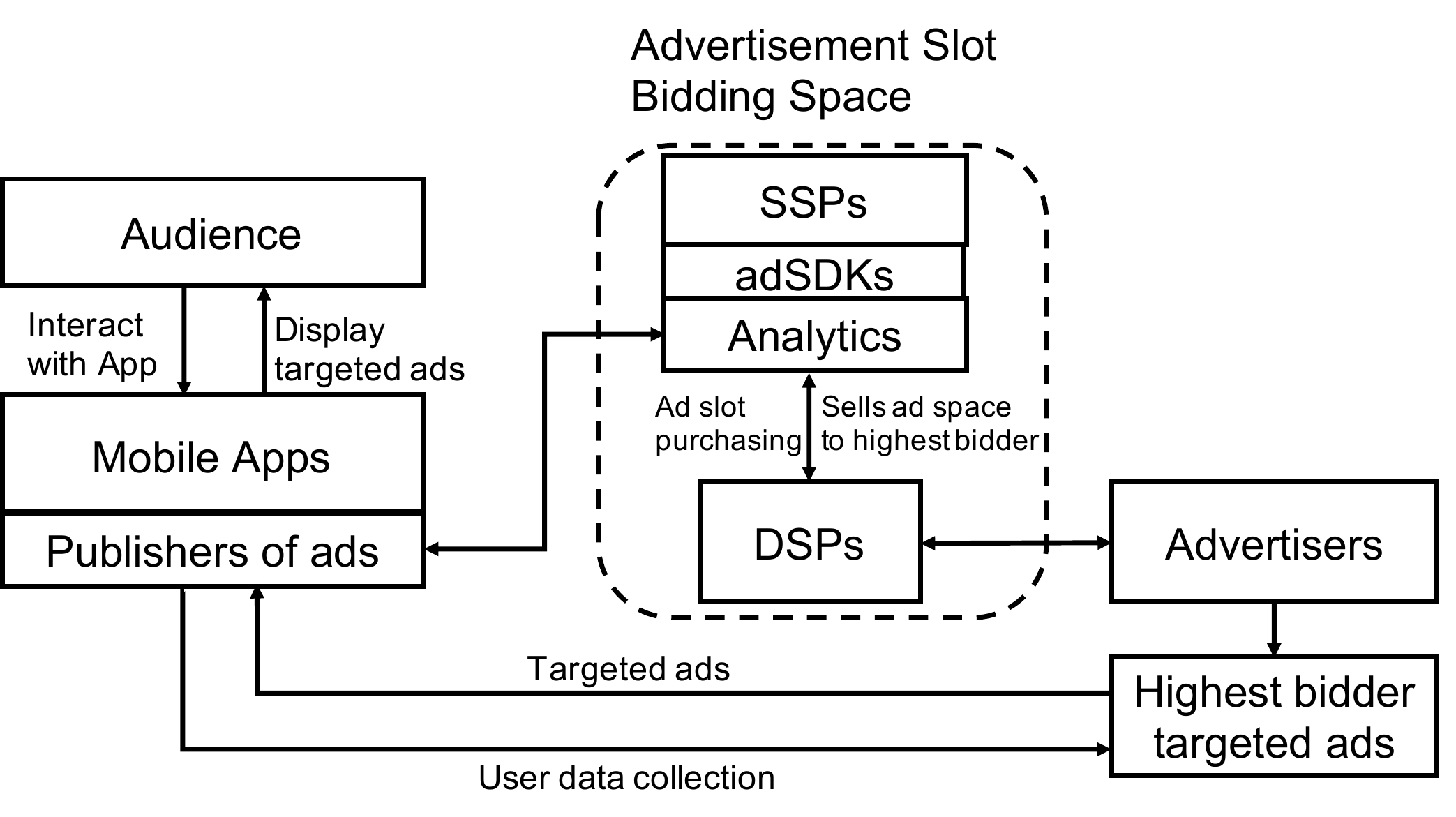}
\caption{Mobile advertisement ecosystem.}
\label{fig:ad_ecosystem}
\vspace{-20pt}
\end{figure}

\subsection{Mobile Advertisement Ecosystem}
The digital advertising ecosystem consists of four types of entities: audiences and publishers, sell-side platforms (SSPs), demand-side platforms (DSPs), and advertisers, as shown in Fig.~\ref{fig:ad_ecosystem}. Audiences are the users who watch the ads when they interact with the contents of a publisher. Publishers are the owners of websites or apps that serve ads, which include SSP toolkits, such as analytic scripts and advertising libraries (i.e., adSDKs for Mobile). DSPs facilitate purchasing ad slots and serving ads on behalf of an advertiser. SSPs facilitate selling the ad spaces to the highest bidder in a publisher's content by auctioning them to DSPs. Advertisers are entities that have ads to display. Advertisers may upload the actual ad content, known as ad creatives, to a DSP, or host them on their servers and provide URLs for the DSP to display.

In the web ad environment, the third party cookie has been the universal tool for tracking host information to provide targeted ads. 
Any website that uses the ad domain can access the cookie of this particular ad domain, which allows for cross-site targeted advertising. 
In contrast, mobile in-app ad environment does not use shared cookies for tracking. Instead, the mobile advertisement ecosystem relies on application stimulus, which collects private data protected by permissions. Our analysis includes a detailed inspection of the tracking data that comes through as macro parameters in the Uniform Resource Locator (URL) of the network communications from mobile devices. 

Apparently, the more information SSPs can provide to the bidders, the higher bids they will get. Therefore, SSPs are motivated to collect a variety of information, such as: mobile advertising identifiers (MAIDs), locations, network profiles, device types, etc.

\subsection{Problem Definition}
The goal of this research is to gain insights into the different privacy leakage behaviors of multiple ad libraries across different apps, organizations, and locations, and evaluate if the cross-application ad libraries can correlate the multiple instances of leaked private information for more precise ad targeting. We combine and analyze traffic from different domains that belong to the same organization to achieve a more accurate estimation of collected information by these organizations.

\begin{table}
\centering
\caption{List of PII categories and types}
\begin{tabular}{|p{2.5cm}|p{5cm}|}
\hline
Unique Identifier    & Advertising ID,  Android ID (device ID), hardware serial, IMEI, IMSI, MAC address   \\ \hline
Personal Information & data of birth (DOB), email address, first and last name, gender                              \\ \hline
Location-related             & GPS location, IP address,  zip code                                      \\ \hline
User Credentials           & username, password \\
\hline
\end{tabular}
\label{tab:PII}
\vspace{-10pt}
\end{table}

\begin{table}[t]
\centering
\caption{Supported location granularity  of top 30 mobile ad networks}
\begin{tabular}{|p{4.2cm}|p{2cm}|}
\hline
 Supported finest location granularity  & \# of ad networks  \\ \hline
Up to country level    & 7      \\ \hline

Up to city and business address level & 15           \\ \hline

Up to zip code level  &  4   \\ \hline

Precise Address level   &  4    \\ \hline

\end{tabular}
\label{tab:citylevel}
\vspace{-10pt}
\end{table}

\textit{Personally identifiable information} (PII) has been defined by NIST in 2010 as ``any information that can be used to distinguish or trace an individual's identity". Such information is often collected by the third-party services or ad networks without users' consent. Leveraging existing studies~\cite{razaghpanah2018apps, ren2018bug, ren2016recon}, we summarize a PII list containing 15 elements. We categorize these private elements into four categories, including: (1) unique identifier, (2) personal information, (3) location information, and (4) user credentials, as listed in Table~\ref{tab:PII}. As shown in Table~\ref{tab:citylevel}, among the top 30 mobile ad networks we surveyed, $23$ ad networks provide fine-grained location-based targeted ads tailored for different cities, zip codes, or precise addresses, within which only $4$ ad networks provide targeted ads for precise addresses. Therefore, our measurement switches across \emph{different cities} for studying location-based mobile tracking. 


\subsection{Threat Model}
We define three main threats that induce users' PII leaks for mobile ad networks.

\noindent \textbf{Threat from an organization with multiple domains.} Popular ad networks usually contain multiple third-party services to aggregate more comprehensive private information from different domains. Thus, the ad networks are able to collect users' private information across multiple apps. The organization-level privacy leakage study is of utmost importance to understand the power of these organizations. 

\noindent \textbf{Threat from adware.} Some app developers may collect sensitive information via ad network libraries or other third-party services either directly or indirectly. It is difficult to tell whether such collection is necessary for the app's functionality. Specifically, adware has been designed to actively collect private information to serve more ads. 

\noindent \textbf{Threat from network eavesdroppers.} Network eavesdroppers may get private information by monitoring the network communications. Some of the private information may be leaked in plaintext via HTTP. In our study, we try to evaluate the severity of such privacy leakage and understand what information an eavesdropper can obtain.

\section{Location-Based Measurement Platform}

\begin{figure*}[!ht]
\centering
\includegraphics[width=5in]{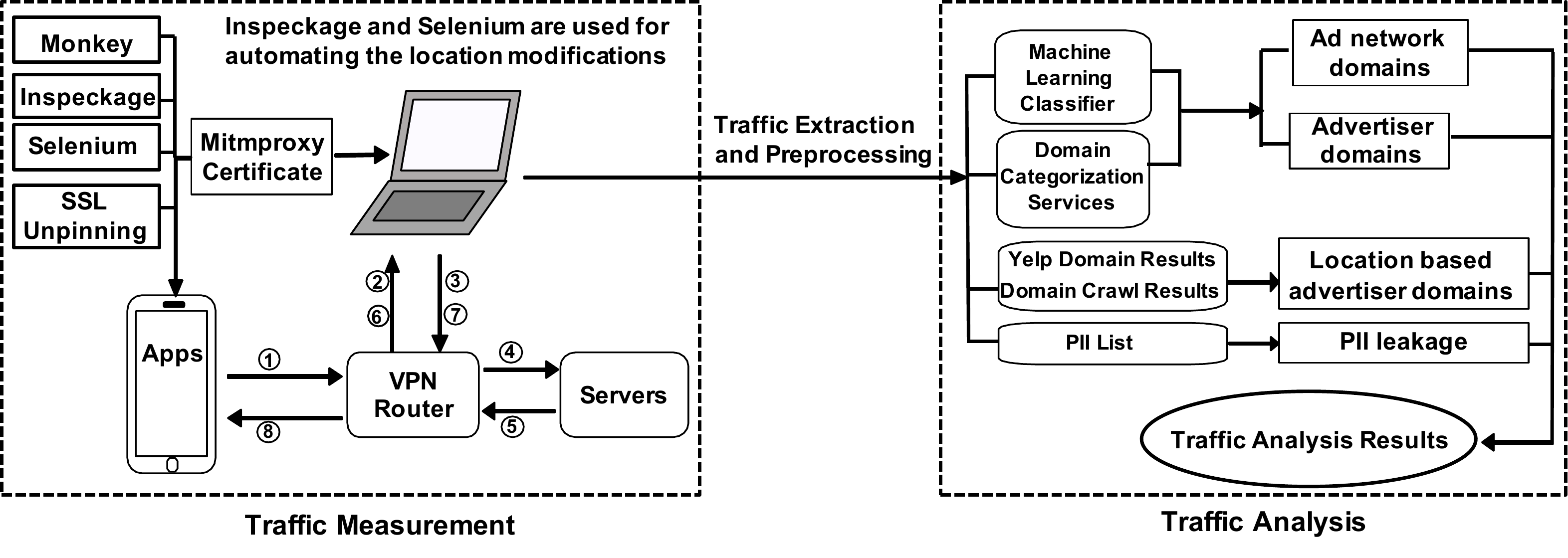}
\vspace{-10pt}
\caption{\textbf{The proposed platform} consists of traffic measurement and traffic analysis. 
Mitmproxy has been set up as a transparent HTTPS proxy. \ding{172} The mobile client initiates a connection to the server. \ding{173} The router redirects the connection to mitmproxy. \ding{174} mitmproxy dynamically generates certificates for the connected hosts and signs it with its own certificate. \ding{175} The mitmproxy connects to the server via router, and establishes a TLS connection. \ding{176} The server with the matched certificate responds to the client. \ding{177} \ding{178} \ding{179} The router will redirect the response to the mitmproxy, and then forward it to the client.}
\label{fig:SystemAtGlance}
\vspace{-20pt}
\end{figure*}

Our measurement platform mainly consists of two components: location-based traffic measurement and traffic analysis, as shown in Fig.~\ref{fig:SystemAtGlance}. 
In-app advertising and mobile web advertising both have their advantages and limitations in the eyes of advertisers. According to eMarketer~\cite{emarketwebandinapp}, mobile apps account for nearly 86\% of time spent using smartphones. But a few top apps dominate the app usage. 
Meanwhile, mobile web advertising may have less usage time, but there are more websites than apps on the market. Some large publisher either do not have apps or their customers tend to use websites more.
Thus, in-app and mobile web advertising are both popular in today's mobile advertising ecosystem, which guide the design of our traffic measurement system.

\subsection{Traffic Measurement}

Our traffic measurement consists of mobile devices, a wireless router, and a workstation. 
Mitmproxy \cite{mitm} is used to intercept the traffic generated by mobile apps. 
We install Mitmproxy certificate on the mobile device to decrypt the HTTPs traffic. We also use Monkey, a popular input generation tool used extensively~\cite{ren2018bug, ren2016recon}, to automate the app interaction by randomly injecting user event sequences. 
We let Monkey interact with each app for five minutes in order to generate enough traffic for analysis. 
Many apps require users to log in with a username and password. To avoid excessive manual efforts, we record and replay the login events using RERAN~\cite{gomez2013reran} for such apps. 

All the traffic between the app and its contacted server would go through the Mitmproxy and the router, where the traffic is intercepted and logged. Mitmproxy is capable of performing TLS interception to record the plaintext of HTTPs requests. For apps that prevent TLS interception using SSL pinning, we use JustTrustMe~\cite{justtrustme} to pass certification verification using SSL unpinning technique. 

The location-based study requires a system to generate genuine location information for large-scale measurement. We use Inspeckage module in Xposed framework to change the locations. To automate the location change, we use Selenium to automatically change the GPS locations through Inspeckage's web interface. Many ad networks cross reference the GPS coordinates with the device's IP address. Therefore, We set up a VPN service using ExpressVPN to fake the IP addresses, which are configured to match the faked GPS locations.

As for the study of mobile web tracking, we aim to identify advertisers engaging in location-based ads. Thus, we query the Yelp Fusion API and select local businesses in different cities whose websites support mobile browsers. We use the proxy service Crawlera to query the websites with fake mobile user agents, and record sites that return no user agent errors.




\subsection{Traffic Analysis} \label{trafficanalysis}
During traffic analysis, we focus on identifying and extracting information from network traffic related to the ad networks. We propose a comprehensive domain classification mechanism to extract the third-party domains, ad network domains, advertise domains, and location-based ad domains.

\noindent \textbf{Third-party domain identification.} 
Domains can be classified as first-party domains and third-party domains, and the owner of first-party domains are the app's owner. We propose an empirical method utilizing domain counts to identify third-party domains, based on the observation that third-party domains appear more frequently across multiple apps than first-party domains. Specifically, for each app in our dataset, we first extract the developer information in the app's webpage. Then, we identify the maximal number of apps that have been developed by the same developer, which we assume will use the same first-party domains. This number is defined as the threshold for identifying the third-party domains. After we extract all the domains in the traffic of all apps, we count the appearance of each unique domain across multiple apps. To avoid bias, we only count once if the domain appears multiple times in each app. 
We identify all the third-party domains, whose number of appearance is higher than the threshold (in our case, 9). As it is possible that potential third-party domains may be counted less than the threshold, we use other methods described below to help catch the missing third-party domains.

\noindent \textbf{Ad network identification.}
We first generate a list of ad network domains using the publicly available information and two \emph{domain organization mapping list} on GitHub~\cite{gitlist1, gitlist2}. This list will be used to identify the ad networks appeared in the collected traffic. There are some unpopular ad networks that are not included in any lists. To identify all the possible ad network domains in our traffic, we utilize the DuckDuckGo search engine to query each domain and get corresponding descriptive information if available. Bi-grams and tri-grams of the descriptive texts are used as their features to classify the domains into ad network domains and non-ad network domains. We construct our training and testing set using 2,000 non-ad network domains from Alexa, and 2,000 randomly picked ad network related domains from EasyList. In the end, the domain classification accuracy reaches around 70\%. 


To improve the classification performance, we propose to use three classification engines, i.e., VirusTotal, McAfee, and OpenDNS, to generate the domain classification result to be our ground truth. These engines are capable of categorizing the domains and quantifying their overall safety. For each unique domain we find in the traffic, we query the classification engines and get the related information such as category, subdomains, and the feedback of Whois lookup for the queried domains. 
If any of the engines considers the domain as an ad network domain, we add it to our ad network domain list.

We evaluate the performance using our ground-truth data with labels, and the domain classification accuracy can reach 92\%. 
Although this list may not cover all the ad networks in the market since these engines cannot recognize all the domains, we consider it to be sufficient for our study. The ad network results contain not only all the popular ad networks in AppBrain~\cite{appbrainad}, but also many small ad networks which have insignificant market shares.


\noindent \textbf{Advertiser domain identification.}
The advertiser domains can also be observed in the traffic served by the ad networks. The advertiser domains are associated with businesses that post ads through ad networks. In order to identify advertisers, we refer to the three popular domain categorization services mentioned above. If any of these three services categorize the domain into advertisements (ad networks), application and software download, web analytics, and other web related categories, we consider them as non-advertisers. We consider all the remaining domains associated with other categories (shopping, education, travel, etc) as advertiser domains, and remove the ones that could not complete the categorization of all three engines. 
To further improve the accuracy of the advertiser list, we utilize Yelp API and query the top 1,000 business domains for each category (if available) at different locations. We add any domains that appeared in our Yelp results to enrich our advertiser domain list.

\noindent \textbf{Location-based ad domain identification.}
After classifying the advertisers' domains, we move on to identify whether the advertiser's related ads are location-based ads. Our goal is to  associate the location of the advertiser domain with the location of the served ads. For example, if the ads appear in the area served by the advertiser, it is considered as location-based.  The service area of the advertiser can be derived  by querying Yelp. But simply relying on Yelp's query results may not be sufficient in identifying the local businesses. To differentiate between the local businesses and all other advertisers, we crawl all the advertiser domains and check if the front page of each domain contains the city name where the served ads appear. By combining the yelp local business list and web crawling result, we can identify the location-based ad domains.

\noindent \textbf{PII leakage identification.}
Mitmproxy provides a standard method of reading and parsing the captured traffic. We use Mitmproxy to extract the information from the traffic flows including the domains and any PIIs. The leakage identification algorithm is presented in Algorithm~\ref{alg:algo1}. We first extract the HTTP/HTTPs request URL, response URL, and request/response contents. By integrating the domain organization mapping lists mentioned above~\cite{gitlist1,gitlist2}, we generate a complete leakage parameter dictionary for every organization. Then, we look up the leakage parameter dictionary to identify the known PIIs values (including hashed values with MD5, SHA1, SHA256, and SHA512) and evaluate the severity of ad networks' PII leakage at different levels including app-level and organization-level across different locations.

\setlength{\textfloatsep}{0.1cm}
\begin{algorithm}[!t]
\scriptsize
\textbf{INPUT:} Predefined PII list (according to Table~\ref{tab:PII}), Domain organization mapping list.\\
\textbf{OUTPUT:} PII leakage of each app
\begin{algorithmic}[1]
\For{\textbf{each} App}
\For {\textbf{each} location}
\State Extract Gets and Posts URLs from captured traffic flows
\State Extract key-value pairs from the URLs
\State Match the key-value pairs with hashed PII values in PII list
 \If {find a match}
   \State Log the key-value pair as a PII leakage for the app  
\State Extract domains associated with the key-value pair
\State Match domains to the domain organization mapping list
  \If {find a match}
   \State Log the key-value pair as a PII leakage for the matched organization
\Else
   \State Log the key-value pair as a PII leakage for ``Others" organization
\EndIf
   \EndIf
\EndFor
\EndFor
\State \textbf{Return} PII leakage results
\end{algorithmic}
 \caption{PII Leakage Identification Algorithm}
 \label{alg:algo1}
\end{algorithm}
\setlength{\floatsep}{0.1cm}

\section{Measurement Results and Analysis}
In this section, we present our measurement results based on extensive experiments. We first compare the mobile web ad tracking and in-app ad tracking behaviors. Then, we expose the organization-level cross-app privacy leakage based on the traffic analysis results. Finally, we study the ad networks' data collection behaviors across different locations, i.e., different cities, rural/urban areas. We use $8$ Moto G4 mobile devices with the Android 4.4.4 (compatible with JustTrustMe) or Android 7.1.2 framework to automatically launch traffic measurement and analysis. For apps that fail to run on Android 4.4.4, we rerun them on Android 7.1.2 without SSL unpinning. 

\subsection{Measurement Dataset}
We have collected two traffic datasets to facilitate the measurement study. Dataset\_1 contains traffic from 1,100 popular apps running at two locations (i.e., Lincoln, Nebraska and New York City), while Dataset\_2 contains the traffic from 110 apps (randomly selected from the 1,100 apps of Dataset\_1) running across 10 different locations, detailed in Table~\ref{tab:PII leakage severity in each location}. Within these two datasets, we removed the apps that fail to generate network traffic in all the locations. In the end, we collect 63.0 GB traffic data: Dataset\_1 contains 814,117 traffic flows from 1,026 apps across 2 locations, and Dataset\_2 contains 535,655 traffic flows from 110 apps across 10 locations.

\subsection{Mobile Web Ad Tracking vs. In-app Ad Tracking}
Mobile web ad tracking allows ad networks to collect users' private information during web browsing activities. We collect the HTTP request/response URLs related to mobile web ad tracking and compare them against in-app ad tracking results. 


\noindent \textbf{Finding 1: mobile web ads and in-app ads contact a similar set of popular third-party domains.} For both types of ad tracking, \emph{googleapis.com} is the most popular third-party domain. Despite such similarities, we also find some third-party domains (especially these ad network domains) only exist in the mobile traffic for in-app ads, such as \emph{flurry.com}, \emph{unity3d.com}, \emph{applovin.com}, \emph{mopub.com}, etc. The reason is that: different from in-app ad tracking that tracks both ad networks' and advertisers' domains, web ad tracking only tracks the advertisers' domains. 

\noindent \textbf{Finding 2: mobile web ads have a significantly lower adoption rate of HTTPs than mobile in-app ads.} 
We also compare the total percentage of HTTP/HTTPs traffic flows originated from domains related to web and in-app location-based ads. As shown in Table~\ref{tab:comparehttpandhttps}, we can see that HTTP traffic dominates in the mobile web ad traffic. The reason is that many landing pages contain third-party HTTP content, which can cause mixed-content errors if the landing pages upgrade to HTTPs. The low adoption rate of
HTTPs in mobile web ads is likely to continue as long as third
parties continue to use HTTP by default.
On the other hand, mobile in-app ads mostly carry HTTPs traffic. The reason is that in-app ads do not use HTTP referrer headers to indicate the sources of the redirected traffic, and thus will not incur mixed-content errors. Without such legacy issue, mobile in-app ads tend to adopt HTTPs for secure app-server communications. 

\begin{table}[h]
\centering
\caption{Comparison of HTTP/HTTPs traffic from web/in-app advertiser domains}
\begin{tabular}{|p{3.5cm}|p{3.5cm}|}
\hline
HTTP traffic (web/in-app) & HTTPs traffic (web/in-app)  \\ \hline
84.8\% / 18.48\%           & 15.2\% / 81.52\%    \\ \hline
\end{tabular}
\label{tab:comparehttpandhttps}
\end{table}


\begin{figure}[!h]
	\centering
	\begin{subfigure}[t]{1.72in}
		\centering
		\includegraphics[width=1.6in]{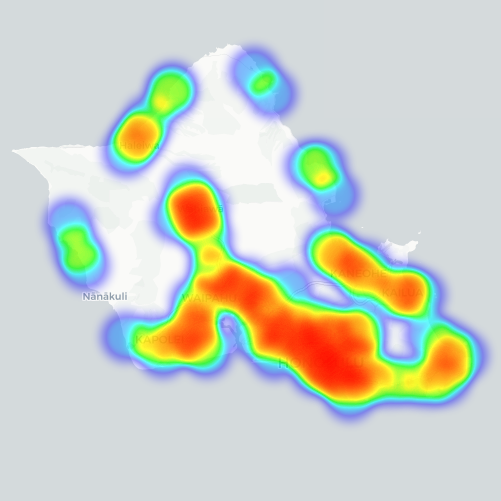}
		\caption{Oahu}\label{fig:1a}		
	\end{subfigure}
	\begin{subfigure}[t]{1.6in}
		\centering
		\includegraphics[width=1.6in]{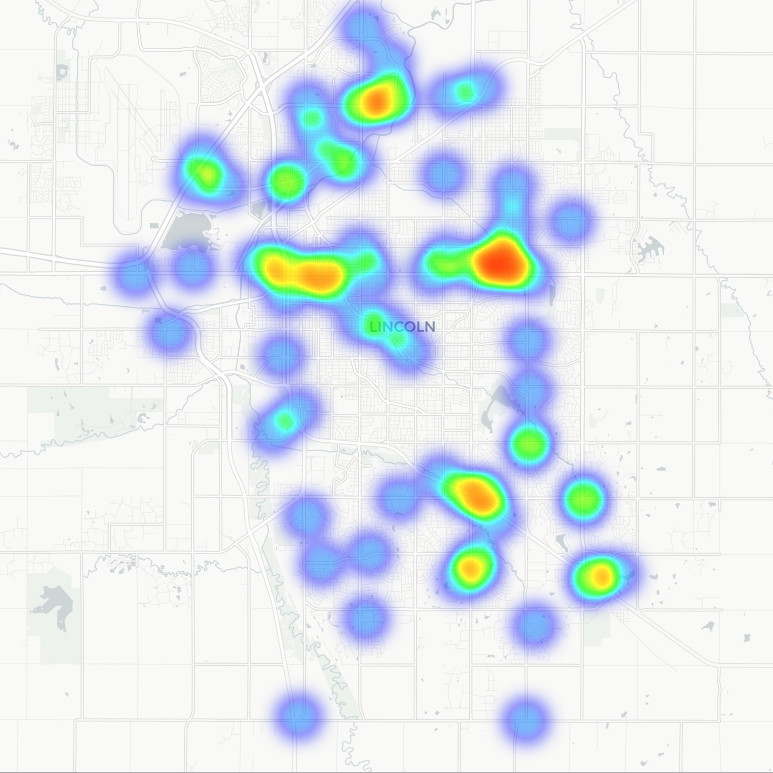}
		\caption{Lincoln}\label{fig:1b}
	\end{subfigure}
	\caption{Heatmap of location requests in landing URLs on the island of Oahu and Lincoln.}\label{fig:heat_map_location_requests_in_landing_urls}
\end{figure}

\noindent \textbf{Finding 3: mobile web ads request location via landing URLs leading to privacy leakage concerns.}
In our web ad traffic, we discover a significant amount of advertisers who seek location information via the landing URLs' macro parameters, without explicitly expressing their purposes. These reckless behaviors allow eavesdroppers to collect and infer sensitive information about users by observing the ad traffic passing through the network. Fig.~\ref{fig:heat_map_location_requests_in_landing_urls} shows the heatmap of location requests (with hot zones and cold zones) in landing URLs of mobile web ads on the island of Oahu and Lincoln. We can see that the location requests tend to be positively related to the population distribution. These observations motivate us to further investigate the privacy leakage through mobile ad networks across different organizations, as well as various locations with an emphasis on the hot zones. 


\subsection{Organization-level Cross-app Privacy Leakage}
The advertising organizations usually own multiple ad networks, and it is conceivable that they tend to aggregate data from these ad networks to achieve a better user profiling for targeted ad delivery.  
In this section, we expose the organization-level data collection behaviors of popular ad networks. We also identify the different organization-level privacy leakage behaviors across different locations.



AppBrain~\cite{appbrainad} 
provides a list of the ad network popularity based on the number of installs of related apps. Similarly, we rank these ad networks based on the collected network traffic of Dataset\_1. 
The result indicates that AdMob (i.e., Google ad network) is observed in the traffic flows of 601 (i.e., 58.58\%) app, demonstrating the wild popularity of Google's ad network. Moreover, Unity 3d, ranked second, is observed in 180 (17.54\%) apps' traffic flows.
This result is consistent with the ad network popularity list of AppBrain website. 


\begin{table}[t]
\centering
\caption{Number of unique domains identified in each category}
\begin{tabular}{|p{1cm}|p{1.4cm}|p{1.2cm}|p{1.3cm}|p{1.8cm}|}
\hline
Dataset & All domains & Top-level domains & Ad domains & Location-based ad domains \\ \hline
Dataset\_1 & 7,208  & 2,532          & 970                & 208                               \\ \hline
Dataset\_2 & 4,398  & 1,760          & 539                & 141                               \\ \hline
\end{tabular}
\label{tab:numberOfDomains} 
\end{table}


Table \ref{tab:numberOfDomains} presents the number of domains identified in each domain category. Within all the domains in Dataset\_1 (or Dataset\_2), we identify 2,532 (or 1,760) unique top-level domains by combining multiple sub-domains. 
These domains belong to 496 (or 247) different organizations. 
Using the domain classification methods in Section~\ref{trafficanalysis}, we can identify 970 (or 539) unique ad domains and 208 (or 141) unique location-based ad domains, respectively. Location-based advertisement constitutes 21.44\% (or 26.16\%) of all the captured advertisements.





\begin{figure}[t]
\centering
\includegraphics[width=1\linewidth]{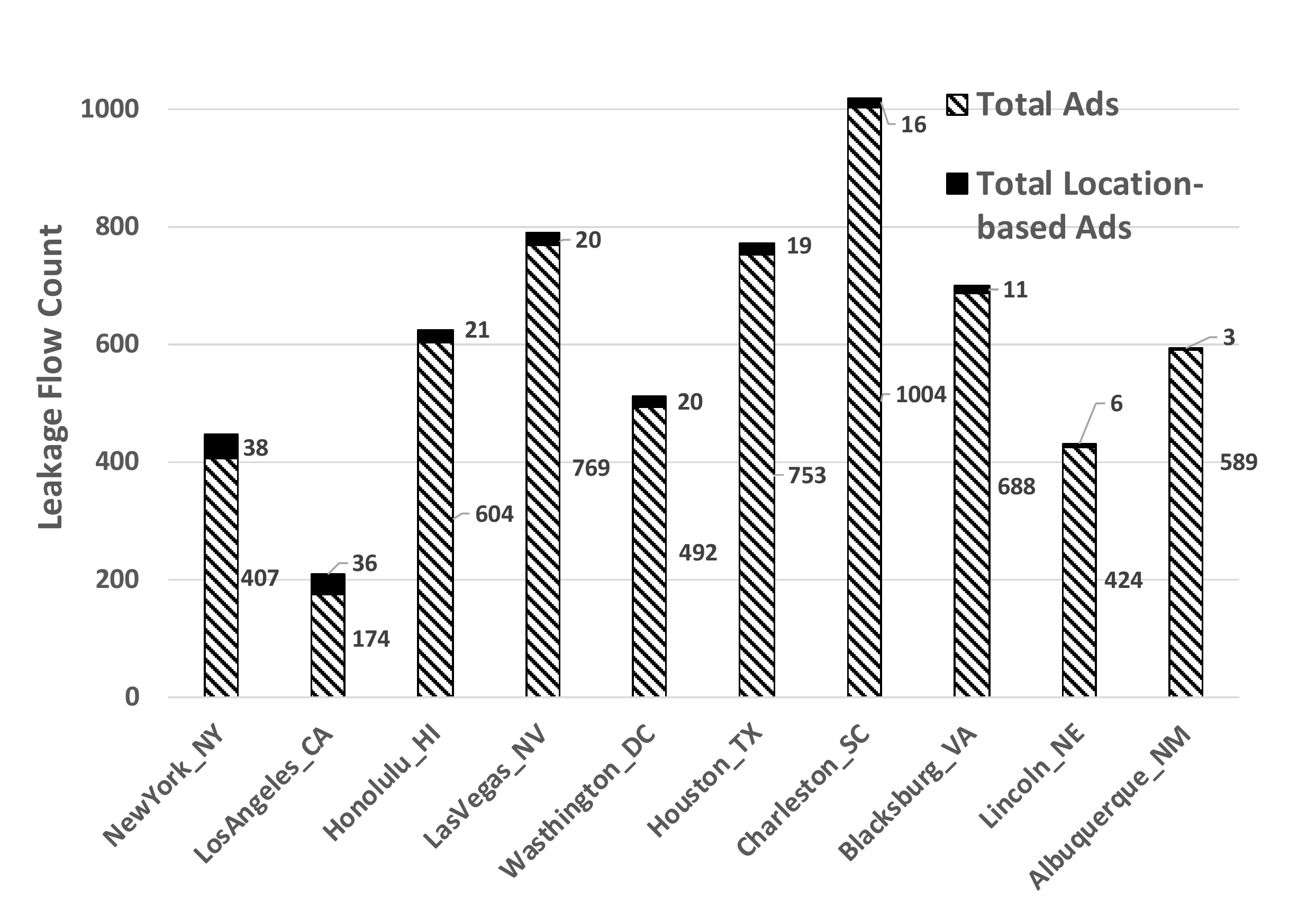}
\caption{Number of total ads and location based ads in each city (Dataset\_2).}
\label{fig:Number_of_Targeted_Ads_domains_in_Each_City_and_Nearby_Cities_10_locations}
\end{figure}

\noindent \textbf{Finding 4: the number of location-based ads across different cities is positively correlated with the population density.} 
Fig.~\ref{fig:Number_of_Targeted_Ads_domains_in_Each_City_and_Nearby_Cities_10_locations} shows the number of location-based ads in each city based on the Dataset\_2. The result shows that the number of location-based ads is positively related with the population density of the cities, which is similar to the phenomenon observed in the mobile web ad marketing environment. A similar trend is also observed with Dataset\_1. In New York, we identified 782 unique advertisers and 173 unique location-based advertisers, while in Lincoln we identified 322 unique advertisers and 72 unique location-based advertisers.

\begin{table}[t]
\centering
\caption{Top 10 ad organization list of PII leakage severity in all locations sorted by average PII leakage flow count per app.}
\begin{tabular}{|p{1.5cm}|p{2cm}|p{3cm}|}
\hline
Organization  &  Average \# of PII Leakage per app   & \# of PII Leakage Types  \\ \hline
LKQD           & 3,399                & 3   \\ \hline
AOL           & 360                & 3                  \\ \hline
Facebook           & 322                & 11                \\ \hline
SpotX          & 279                & 6                  \\ \hline
Tapjoy           & 211                & 9                 \\ \hline
Heyzap           & 184                & 6                 \\ \hline
Google           & 80                & 15                  \\ \hline
AppsFlyer           & 70                & 7                 \\ \hline
MoPub           & 54                & 2                 \\ \hline
Applovin           & 37                & 2                   \\ \hline

\end{tabular}
\label{tab:orgall} 
\vspace{-5pt}
\end{table}


We examine the app-level privacy leakage and find that Game apps are leaking private information over a large number of flows, and they leak different types of PII information. 
These top-ranked apps all interact with multiple ad networks (i.e., $8$ ad networks on average), and the organizations behind these ad networks are able to aggregate a considerable amount of private information.

\noindent \textbf{Finding 5: Popular ad networks generally collect more diverse types of PII data.} 
For data aggregation at the organization level, Table~\ref{tab:orgall} shows the top 10 list of the ad organizations ranked by the PII leakage severity in terms of average leakage flow counts per app (i.e., total leakage flow counts of an ad network divided by the number of apps associated with this ad network) across all the locations. In general, the result indicates that some popular ad networks (e.g., Facebook) generate a large amount of PII leakage flows per app.
A considerable number of flows from LKQD, a video ad platform (which is included in multiple apps, such as \emph{cjvg.santabiblia} and \emph{com.july.ndtv}), leak private information, although it only leaks three different types of PII information. The top 5 ad organizations ranked by the number of unique PII leakage types are: Google, Facebook, Amazon, ironSource, and Tapjoy. This indicates that the big companies with popular ad networks collect most types of privacy information. It is worth noting that Amazon collects 11 types of PII information, while we only find 736 flows carrying private information that are associated with Amazon ad domain within our datasets.



\begin{figure*}[!ht]
\centering
\includegraphics[width=5in]{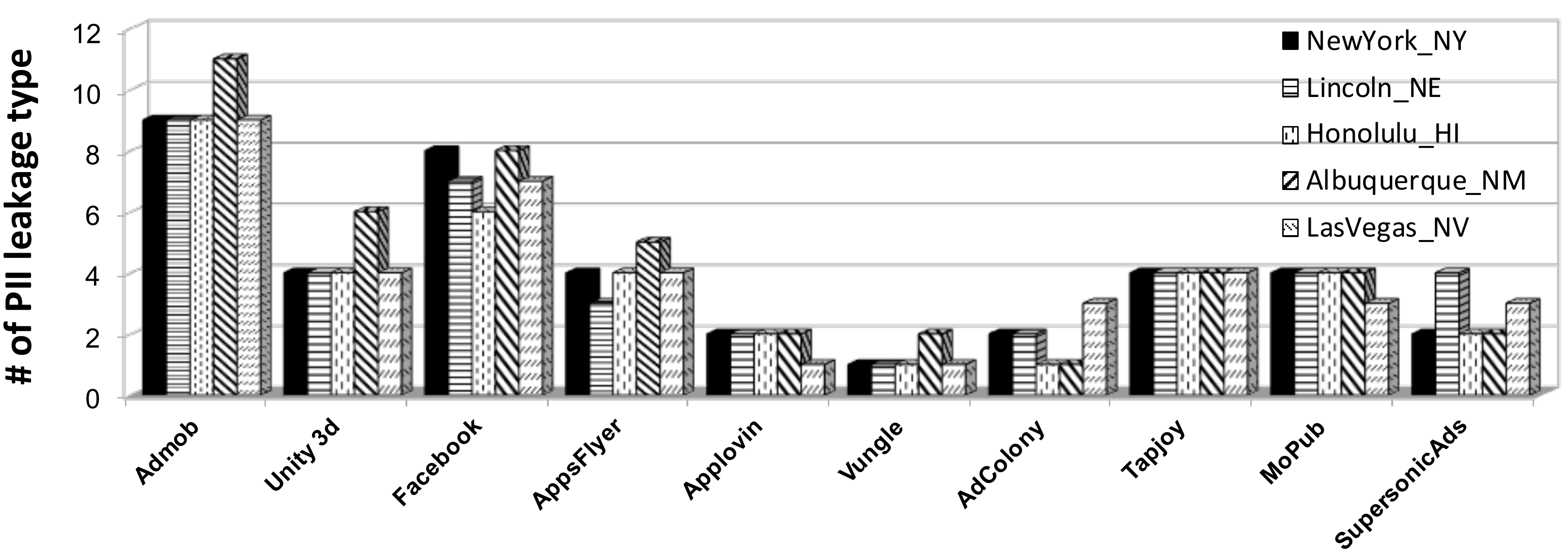}
\vspace{-5pt}
\caption{The leaking PII types of top 10 ad networks across different locations.}
\label{fig:5location10networks}
\vspace{-15pt}
\end{figure*}

\begin{table}[h]
\centering
\caption{Location related privacy leakage observed in Dataset\_1}
\begin{tabular}{|p{2cm}|p{1cm}|p{1.5cm}|p{1cm}|p{1cm}|}
\hline
City name  &  GPS & IP address & Zip code     & Total                \\ \hline
New York           & 32,826   & 29,161  & 5,857     & 67,844             \\ \hline
Lincoln           & 25,761     & 20,765  & 4,691    & 51,217             \\ \hline
\end{tabular}
\label{tab:locationdataset1}
\vspace{-10pt}
\end{table}



\begin{table}[h]
\centering
\vspace{-10pt}
\caption{The mean and standard deviation (STD) of PII leakage flow across 10 locations}
\begin{tabular}{|p{2.5cm}|p{2.5cm}|p{2.5cm}|}
\hline
PII Type&Mean&STD\\ \hline
Advertising ID&7,762.22&2,311.37\\ \hline
IP address&1,585.85&922.73\\ \hline
GPS&2,072.31&804.21\\ \hline
MAC address&778.83&751.83\\ \hline
Android ID&1,214.82&245.68\\ \hline
email&407.72&103.26\\ \hline
gender&112.69&50.59\\ \hline
IMEI&93.35&46.13\\ \hline
Hardware serial&13.17&4.95\\ \hline

\end{tabular}
\label{tab:PII leak mean and sdiv}
\vspace{-10pt}
\end{table}

\begin{table}[h]
\centering
\vspace{-10pt}
\caption{PII leakage severity at each location.}
\begin{tabular}{|p{2.5cm}|p{2cm}|p{3cm}|}
\hline
City name  & \# of PII leakage & Ad networks with maximal collected PII\\ \hline
Las Vegas           & 22,328                &  AdMob                 \\ \hline
Albuquerque           & 17,699                &  LKQD                 \\ \hline
Honolulu           & 16,996                &   LKQD                \\ \hline
Washington, D.C.           & 16,005                &    LKQD               \\ \hline
Charleston           & 14,576                &  AdMob                \\ \hline
Blacksburg           & 14,069                &   AdMob                \\ \hline
Houston           & 13,095                &       AdMob            \\ \hline
Los Angeles           & 11,875                &     LKQD              \\ \hline
Lincoln           & 10,808                &       AdMob            \\ \hline
New York           & 10,140                &       AdMob            \\ \hline
\end{tabular}
\label{tab:PII leakage severity in each location}
\vspace{-10pt}
\end{table}

\begin{table}[h]
\centering
\vspace{-10pt}
\caption{The PII type that is most frequently leaked by the ad networks based on 2 datasets.}
\begin{tabular}{|p{2.5cm}|p{2.5cm}|p{2.5cm}|}
\hline
PII Type&Ad Network&Collected Times\\ \hline
Advertising ID&LKQD&72,185\\ \hline
IP Address&LKQD&34,584\\ \hline
GPS&LKQD&28976\\ \hline
MAC address&Tapjoy& 5,364\\ \hline
Android ID&Tapjoy&7,690\\ \hline
email&Google&4,403\\ \hline
gender&Appodealx&343\\ \hline
IMEI&Fyber&87\\ \hline
Hardware Series&Charboost&27\\ \hline
\end{tabular}
\label{tab:PII leaked most ad network}
\end{table}

\subsection{Location-based Private Data Collection of Ad Networks}

Ad networks extensively collect users' location information. Table~\ref{tab:locationdataset1} shows that the ad networks collect location information in the format of GPS, IP address, and zip code. We observe that New York has more location-related leakage compared with Lincoln. This result complies with our assumption that ad networks in larger cities will initiate more location related requests and collect more location data. The experiment with Dataset\_2 presents similar phenomenon, which we omit here due to page limitation. 

Before we unveil the details of ad network collection behaviors across different locations, we evaluate the difference among the leaked PIIs  across different locations. 
Table~\ref{tab:PII leak mean and sdiv} shows the mean and standard deviation for the number of PII leakage flows of each PII type to measure the magnitude of the differences across 10 locations. From this table, we can see that the number of PII collections varies significantly across locations, while the Advertising ID, IP address, and GPS location are the most collected PII types for these mobile ad networks. This observation indicates that the ad networks behave differently in collecting users' private information across different locations. 

\noindent \textbf{Finding 6: The number of ad networks' PII leakage flows differs across different cities.} 
To further identify the private data collection behaviors of ad networks across different locations, we extract the traffic flows related to the ad domains, measure the total number of PII leakage flows and the number of PII leakage types at each location. Table~\ref{tab:PII leakage severity in each location} shows 
the number of PII leakage flows vary across different locations. In addition, AdMob collects the maximal number of privacy-leaking flows within $6$ cities. It is worth noting that AdMob collects the most privacy-leaking flows in almost all cities, while LKQD collects the most privacy-leaking flows in 4 cities, but it keeps quiet (i.e., collects negligible amount of privacy-leaking flows) in other cities, maybe due to its failure in the ad space bidding in these cities. 
Fig.~\ref{fig:5location10networks} shows the different number of PII leakage types of ad networks across different locations. Overall, AdMob collects the most types of PIIs across all locations. 


These ad networks present different behaviors across different locations, and we suspect that different ad networks may be interested in different PII types. In Table \ref{tab:PII leaked most ad network}, we show the number of times that each ad network collects the corresponding PII information. We show the ad network with the maximal collection times, which indicates that the ad network is most interested in the corresponding PII. LKQD has the most interests in the Advertising ID, IP address, and GPS, while Google is most interested in email address.

We examine the privacy policy of all the ad networks, and we find that all the ad networks claim to collect both fine-grained location (GPS) data and coarse-grained location (IP address) data, which we have confirmed using our measurement study. Even though all the ad networks claim to collect both fine-grained and coarse-grained location data, they are still different from each other in terms of the number of decimals in the collected GPS location data. To put it into context, when the decimals of GPS data are 3 digits, it can be used to identify the neighborhood or street which is precise to 111.32 meters at the equator. Moreover, when it reaches to 6 digits, it can be used to identify the individuals with the precision of 111.21 millimeters at the equator. 



\vspace{-10pt}
\begin{table}[h]
\centering
\caption{The ad networks' location leakage severity.}
\begin{tabular}{|p{4cm}|p{2cm}|p{1.5cm}|}
\hline
Ad Networks & \# of decimals in collected GPS  & \# of valid decimals \\ \hline
Sitescout, Mopub, Google, Appodeal, mediabrix.com, adsrvr.org, Amazon            & 15                     & 8              \\ \hline
Nexage, algovid.com,
adhigh.net, LKQD, fqtag.com,
AdColony            & 14                     & 8              \\ \hline
xAd, Flashtalking            & 13                     & 8              \\ \hline
Facebook, advertising.com           & 8                     & 8              \\ \hline
1rx.io, PubMatic            & 7                     & 7              \\ \hline
OpenX, Yandex, Inneractive, SpotX, Casale Media, Unity 3D, smartadserver.com, StreamRail, Smaato            & 6                     & 6              \\ \hline
Vungle, AdBuddiz, Heyzap, Applovin            & 3                     & 3              \\ \hline
Adform, Millenial Media, InMobi            & 1                     & 1              \\ \hline
\end{tabular}
\label{tab:All Ad networks Location Leakage Severity} 
\vspace{-10pt}
\end{table}

\noindent \textbf{Finding 7: most ad networks collect fine-grained GPS location data.} 
Table \ref{tab:All Ad networks Location Leakage Severity} presents all the ad network location leakage severity. Among the 35 ad networks, 28 of them have collected user's fine-grained location (i.e., the number of decimals is greater than or equal to 6). We consider these ad networks to be aggressive in collecting precise locations since they have the ability to locate individuals very precisely. As a result, these ad networks can provide advertisers with a precise location targeting service. Moreover, among these ad networks, 7 of them collect location data with the 15 digits decimal accuracy, 
indicates at least 7 digits in the GPS data are useless. 
Potential malicious ad network or attacker can take advantage of these extra digits to embed some users' sensitive information and send to the server without getting spotted.

\subsection{Rural Area vs. Urban Area Location-based Mobile Tracking}

The aforementioned experiments prove the different  tracking behaviors of mobile ad networks across different cities. As shown in Table~\ref{tab:citylevel}, most ad networks support location-based ads with respect to different cities. We set up an experiment to verify whether these ad networks behave the same at rural area and urban area in the same city. We select $10$ popular apps, which collectively include $29$ ad libraries. 
We also pick two locations, i.e., the \emph{Downtown} and \emph{Lake Ray Roberts}, in a large city \emph{Dallas} for comparison. We further randomly pick 5 points in the Downtown area, and 5 points in the Lake area. To avoid the time variability, two comparative tests (i.e., one in Downtown and one in Lake) are performed simultaneously using the same apps with the same recorded user inputs. We run each app for 10 minutes, test it 10 times at each point, and collect the network traffic. 

\noindent \textbf{Finding 8: A more diverse group of mobile ad networks operate in rural area, which results in more PII leakage.} We notice similar number of PII leakage for most PII types at these two different locations, while significant difference can be observed for four PII types as shown in Table~\ref{tab:ruralurban}. Generally, the rural area (i.e., Lake) collects more PII data than the urban area (i.e., Downtown), which is counterintuitive. Delving into the traffic, we notice a more diverse group of ad networks in the Lake area, who collect more PII information. 
Notably, LKQD and Tappx collect most of the PII information in the Lake area, but both never show up in the Downtown area. This can be attributed to the less competition in the Lake area for the ad bidding system, which brings in ``less competitive " players in mobile ad business. On the other hand, the Downtown area is a highly competitive area for mobile ad networks, where ``more competitive" players win with a high probability.

\begin{table}[h]
\centering
\caption{Location related privacy leakage for rural/urban area}
\begin{tabular}{|p{1.3cm}|p{1.2cm}|p{1.3cm}|p{1.5cm}|p{1.5cm}|}
\hline
Location  &  Average \# of GPS & Average \# of IP & Average \# of Ad ID & Average \# of Android ID\\ \hline
Downtown    & 10  &  42 &   35  &    22        \\ \hline
Lake           &  47 & 144 &   131  &  210            \\ \hline
\end{tabular}
\label{tab:ruralurban}
\vspace{-10pt}
\end{table}

\section{Discussions and Future Work}
\noindent \textbf{SSL pinning and input automation.} We investigate the private data collection behaviors of ad networks across different locations. We use real devices for our measurement study to avoid the emulation detection of some sophisticated apps. Higher version of Android system has implemented a stricter rule in preventing SSL unpinning, in which the developers can prevent traffic interception by trusting only specific/allowed certificates. As a result, we cannot decrypt HTTPs traffic from several apps which only work with high Android version. Also, SSL unpinning does not work with all the apps, This is a common limitation of traffic analysis on Android devices. 
Our results are based on the traffic of the apps that could be captured. 
We also use Monkey to automate the user input generation, and the proposed study will benefit from the advancement of input generation tools~\cite{wong2016intellidroid} to improve the coverage.

\noindent \textbf{App execution time.} Our automated platform only executes one app for $5$ or $10$ minutes. However, in location-based advertising, the app's execution time can be a key element that impacts the traffic collection results. Some advertisers prefer to provide their location-based ads during a specific time of the day so that they can maximize the effectiveness of their ad delivery. For future study, we will record all the timestamps of our traffic and find out which period of time is the ``golden" collection time for different ad networks, and how these ad networks' behaviors will change at a different time. 

\noindent \textbf{Traffic obfuscation.} Obfuscation has been used by apps to encrypt users' private data like username, password, or email. As mentioned in Section~\ref{trafficanalysis}, we  hash all the known PII values with different hash functions to match traffic. However, if the malicious ad networks or attackers intentionally try to evade our analysis, they can steal the users' PII without getting spotted by using customized hash functions or encryptions.

\section{Conclusion}

In this paper, we present a measurement study for privacy leakage in location-based mobile advertising service. We proposed and implemented a transverse measurement platform for mobile ad networks capable of location spoofing, domain classification, and privacy leakage detection. We performed extensive threat measurements and assessments with the collected traffic data. 
Our findings show that mobile web tracking and in-app tracking share a similar set of third-party domains, and the exceedingly high percentage of HTTP requests in mobile web ads becomes a vulnerable point inciting eavesdropping attacks. 
Our results verified that ad networks perform differently across different locations, and most ad networks can extract precise locations. Alarmingly, there is little correlation between ad network size and their location information leakage severity since both large and small ad networks could collect or infer fine-grained location information.

\section*{Acknowledgements}

We would like to thank the anonymous reviewers for their valuable
comments and feedback. This work was supported in part by  the NSF grants CNS-1566388, CNS-1717898, CNS-1566443, DGE-1662487, and Louisiana Board of Regents under grant LEQSF(2015-18)-RD-A-11.



%



\bibliographystyle{IEEEtran}
\bibliography{references}

\end{document}